\documentstyle[preprint,eqsecnum,aps]{revtex}
\begin{document}
\draft
\preprint{IU/NTC 97-07,~~ TIFR/TH/97-38}
\title{
Using Drell-Yan Processes to Probe Nucleon and \\
Meson Structure Functions
}
\author{R.S. Bhalerao \cite{email1}}
\address{
Theoretical Physics Group \\
Tata Institute of Fundamental Research \\
Homi Bhabha Road, Colaba, Mumbai 400 005 INDIA}
\author{J.T. Londergan \cite{email2}}
\address{
Department of Physics and Nuclear Theory Center \\
Indiana University \\
Bloomington IN 47408 USA}
\maketitle
\begin{abstract}
We investigate how Drell-Yan processes can be used to measure the
magnitude of flavor symmetry violation in the proton sea. We examine
the utility of the following beams: protons, charged pions, and
charged kaons. In each case we present an approximate expression for
the Drell-Yan asymmetry. Using currently available parton
distributions, we locate those kinematic regions which provide the
greatest information on the quantity $\bar{d}^p(x) - \bar{u}^p(x)$. If
sufficiently intense $K^+$ beams were available, they could provide an
efficient measurement of this quantity. Finally we present and discuss
sets of sum rules for the Drell-Yan processes.
\end{abstract}

\narrowtext

\section{Introduction}

At the present time there is considerable interest in the flavor
structure of the nucleon sea \cite{kum}. This is largely a consequence
of recent experimental results, such as the discovery by the New Muon
Collaboration (NMC) of a violation of the Gottfried sum rule
\cite{nmc}. The most likely explanation of the NMC measurement was a
significant flavor asymmetry in the proton sea, i.e.~ $\bar{d}^p (x)
\ne \bar{u}^p (x)$. Although there is no compelling reason why one
should have $\bar{d}^p(x) = \bar{u}^p(x)$, and in fact due to the
Pauli principle one expects $\int_0^1 dx\,\left[ \bar{d}^p(x) -
\bar{u}^p(x)\right] > 0$, phenomenological fits to parton 
distributions prior to the NMC measurement had typically set these sea
distributions equal.

Ellis and Stirling \cite{es} pointed out that this asymmetry could be
measured in proton-induced Drell-Yan (DY) processes on hydrogen and
deuterium targets. They advocated measurements taken at center-of-mass
rapidity $y=0$, i.e.\ $x_1 = x_2$, where $x_1$ ($x_2$) is
approximately the momentum fraction carried by the projectile (target)
quark or antiquark which annihilates in the DY process. These
measurements were carried out by the NA51 collaboration at CERN
\cite{na51}, which obtained the value
\begin{equation} 
{\bar{u}^p(x) \over \bar{d}^p(x)} \Big | _{<x> = 0.18} = 
0.51 \pm 0.04 (stat) \pm 0.05 (syst).
\end{equation}

The $\bar{u}^p/\bar{d}^p$ distributions have recently
been measured
in the E866 experiment at FNAL \cite{e866}, through proton-induced DY
processes on hydrogen and deuterium. The E866 group has measured
$\bar{u}^p/\bar{d}^p$ over a range of $x_1$ and $x_2$ values. With the
Main Injector upgrade scheduled for completion in early 1999 at FNAL,
there is a revival of interest in the physics issues which could be
addressed with the beams of kaons that could become available. We
think that there is interesting physics which could be done in that
program.

In this paper we will review the prospects for measuring the flavor
asymmetry in the nucleon sea through DY processes induced by protons
or charged mesons ($K^{\pm}$ and $\pi^{\pm}$). We will examine the
sensitivity of various cross sections to the up/down antiquark ratio
in the proton. We will also look at the sensitivity to the kinematic
region for both projectile and target. We will discuss what
measurements are needed if the DY process is to be used to pin down
the largely unknown kaon structure functions. Finally we will present
and discuss sets of sum rules for the DY processes.

\section{Analytic Approximations for Drell-Yan Asymmetry}

The general form of the DY cross section is:
\begin{eqnarray}
 {d^2\sigma^{DY}\over dx_1\,dx_2} &=& \left({4\pi \alpha^2\over 9 M^2}
  \right) \sum_{j} e_j^2 \left[ q_j^B(x_1)\bar{q}^T_j(x_2) + 
   \bar{q}_j^B(x_1)q^T_j(x_2) \right] \nonumber \\ 
   &\equiv& \left({4\pi \alpha^2\over 9 M^2} \right) H^{DY}(x_1,x_2).
\label{DYdef}
\end{eqnarray}
In Eq. (\ref{DYdef}), $\alpha$ is the fine-structure constant, $M^2 =
s x_1 x_2$ is the dilepton mass squared, $e_j$ is the charge of parton
of flavor $j$, $q$ and $\bar q$ are the quark and antiquark
distributions, and the superscripts $B$ and $T$ refer to the beam and
target, respectively.
  
In order to probe differences between up and down antiquarks in the
proton, one measures the ratio of DY cross sections on hydrogen and
deuterium \cite{na51}. If one makes the impulse approximation, the DY
cross section on deuterium ($D$) will just equal the sum of DY cross
sections on the free proton ($p$) and neutron ($n$). For a beam
particle $B$ the asymmetry $A^{DY}_B \equiv 2 \sigma_{Bp} /
\sigma_{BD} -1$ then becomes
\begin{eqnarray}
A^{DY}_{B}(x_1,x_2) & = & {H_{Bp}^{DY}(x_1,x_2)- 
  H_{Bn}^{DY}(x_1,x_2)\over H_{Bp}^{DY}(x_1,x_2)
  + H_{Bn}^{DY}(x_1,x_2)}.
\label{eq:APp}
\end{eqnarray}
Because $A^{DY}_B$ is a ratio of cross sections, one expects some
systematic errors to cancel. In subsequent sections, we will suppress
for simplicity the dependence of $A^{DY}_B$ on $x_1$ and $x_2$.

In comparing DY cross sections on protons and neutrons, we invoke
charge symmetry for the nucleon parton distributions. This involves,
for example, setting
\begin{equation}
u^n(x) = d^p(x) ~~{\rm and}~~ d^n(x) = u^p(x),
\label{eq:csdef}
\end{equation}
and similarly for the antiquark distributions. We can
therefore define all nucleon parton distributions in terms of those in
the proton. Charge symmetry violating [CSV] effects have been
estimated by Londergan, Thomas and collaborators \cite{rtl,lglrt}.
For these processes the CSV effects should never be greater than 1\%
of the relevant amplitudes.

Contributions to $A^{DY}_B$ which involve an up quark from the
projectile will then be directly proportional to $\bar{u}^p(x) -
\bar{d}^p(x)$. Since up quark terms will be four times as large as 
down quark terms, due to the square of the quark charge, such terms
should make substantial contributions to the measured asymmetries. For
this reason $A^{DY}_B$ should be an excellent place to measure up/down
differences in the proton sea. This does not mean that all projectiles
would be equally good probes of $\bar{u}^p(x) - \bar{d}^p(x)$. In fact
$p$ and $K^+$ turn out to be better than $K^-, \ \pi^+$ and $\pi^-$.
Reasons for this are discussed in section IV.

In subsequent sections, we make the following approximations. We
neglect nuclear corrections to free structure functions in deuterium
\cite{mt1}. We neglect heavy-quark (charm and heavier) contributions. 
We assume $q_{sea}(x)=\bar{q}_{sea}(x)$. Finally, while displaying
various equations we ignore sea-sea contributions; these will be
included in actual calculations, but are neglected here for
simplicity.

\subsection{Drell-Yan Asymmetry for Proton Beams}
  
For proton projectile we have
\begin{equation}
H_{pp}^{DY}(x_1,x_2) = {1 \over 9} \left[4\left( 
  u^p_v(x_1)\,\bar{u}^p_s(x_2) + \bar{u}^p_s(x_1)u^p_v(x_2) \right) + 
  d^p_v(x_1)\bar{d}^p_s(x_2) + \bar{d}^p_s(x_1)d^p_v(x_2) \right],
\label{eq:pp}
\end{equation}
and a similar expression for $H^{DY}_{pn}(x_1,x_2)$. The asymmetry is
\begin{eqnarray}
A^{DY}_{p} & = & {H_{pp}^{DY}- H_{pn}^{DY}\over 
  H_{pp}^{DY}+ H_{pn}^{DY}}  \nonumber \\
  &=& {-\left( 4u^p_v(x_1)- d^p_v(x_1)\right)\left( \bar{d}^p_s(x_2) - 
  \bar{u}^p_s(x_2) \right) + \left( 4\bar{u}^p_s(x_1) - \bar{d}^p_s(x_1) 
  \right)\left(u^p_v(x_2) - d^p_v(x_2)\right)\over 
  \left(4u^p_v(x_1) + d^p_v(x_1)\right) \left( \bar{d}^p_s(x_2) + 
  \bar{u}^p_s(x_2) \right) + \left( 4\bar{u}^p_s(x_1) + \bar{d}^p_s(x_1) 
  \right)\left(u^p_v(x_2) + d^p_v(x_2) \right) }. 
\label{eq:Ap}
\end{eqnarray}

In Fig.\ \ref{fig1}, we show the asymmetry $A^{DY}_{p}(x_1=x_2)$ vs.\
$x_1$. Recall that $x_1 \equiv x_{beam}$ and $x_2 \equiv x_{target}$.
The curves are obtained with the phenomenological nucleon parton
distributions of Gl\"uck, Reya and Vogt \cite{grv} with $Q^2 = 20$
GeV$^2$. (Calculations were also performed with CTEQ-3M \cite{lai} and
MRS(G) \cite{mrsg} parton distributions; results were similar and are
not shown.) The lower curve is obtained with the full parton
distributions, while the upper one results from fixing $\bar{u}^p(x) =
\bar{d}^p(x)$. For this kinematics, the two asymmetries have opposite
signs: If we allow flavor asymmetry, the predicted DY asymmetry is
negative, while it is positive if we require SU(2) flavor symmetry. In
some kinematic regions the asymmetry $A^{DY}_p$ is large, but the DY
cross sections may be so small that it cannot be accurately measured.

Figure\ \ref{fig1} requires that we measure the DY asymmetry along the
line $x_1=x_2$. We can examine whether this is the most sensitive
region to test flavor asymmetry in the nucleon sea. In Fig.\
\ref{fig2}, we show how the asymmetry $A^{DY}_{p}$ varies with $x_1$
and $x_2$. In Fig.\ \ref{fig2}a, we plot equal-$A^{DY}_{p}$
contours. Here and in other figures, the two extreme contours are
labelled; the intermediate contours are understood to have labels
differing in steps of $0.1$. In Fig.\ \ref{fig2}b, we plot the
equal-$A_p^{DY}$ contours obtained by fixing $\bar{u}^p(x) =
\bar{d}^p(x)$. In Fig.\ \ref{fig2}c we show contours of equal
asymmetry difference $\Delta^{DY}_p$ defined by
\begin{equation}
\Delta^{DY}_p(x_1, x_2) \equiv A^{DY}_p |_{\bar{u} =\bar{d}} - 
A^{DY}_p |_{\bar{u} \ne \bar{d}}~.
\label{diff1}
\end{equation}
Along the line $x_1=x_2$ (the dashed curve in Fig. 2a), we see that 
the asymmetry
$A^{DY}_p$ is negative and becomes more negative with increasing $x$.
Secondly, if we set $\bar{u}^p(x) = \bar{d}^p(x)$ and $x_1 = x_2$, the
resulting asymmetry is positive and roughly constant at $A^{DY}_p
\approx 0.1-0.2$; see Fig. 2b. Finally, the difference between the two
asymmetries goes on increasing as $x_1 = x_2$ increases; see Fig. 2c.
All these features are consistent with Fig. 1. Thus the equal-$x$
kinematics is quite adequate if the purpose is to maximize the
difference $\Delta^{DY}_p$.

{\it However, if the purpose is to maximize the asymmetry $A^{DY}_p$,
the equal-$x$ kinematics does not seem to be the optimum choice. The
asymmetry can be significantly altered by deviating from the dotted
line in Fig. 2a.} If $x_1 = x_2 = 0.2$, the asymmetry is about $-10\%$
which is consistent with the experimental observation in
Ref. \cite{na51}. However, if the measurements had been made at $x_1 =
0.3$ and $x_2 = 0.2$, the asymmetry would have been $\approx -20\%$.
On the other hand, if the measurements had been made at $x_1 = 0.1$
and $x_2 = 0.2$, the asymmetry would have been nearly zero.
Experimentally the ratio $(\sigma_{BD}/2\sigma_{Bp})$ is measured, and
then the asymmetry $A^{DY}_p$ is deduced. Contours for a fixed
$(\sigma_{BD}/2\sigma_{Bp})$ are identical to those for a fixed
$A^{DY}_p$ (Fig. 2a); only the contours labels are different.

The E866 group at FNAL \cite{e866} has measured DY processes for
proton projectiles on protons and deuterons over a wide kinematic
region.  They focus especially on the region of large $x_1 \equiv x_{beam}$ 
and small $x_2 \equiv x_{target}$.  For sufficiently large $x_1$, 
the contribution from the projectile sea in Eq.\ \ref{eq:pp} is
negligible relative to the valence contribution.  Since 
$d^p_v(x_1)/u^p_v(x_1) < 1$, in this kinematic region the ratio of 
Drell-Yan cross sections is approximately given by 
\begin{equation}
 {\sigma_{pD}\over 2\sigma_{pp}} = \left[ 1+ A^{DY}_p \right]^{-1} 
 \approx {1\over 2} \left[ 1 + {\overline{d}^p(x_2)\over 
 \overline{u}^p(x_2) } \right] ~~,
\label{eq:DYrat}
\end{equation}
as can be seen from Eq.\ \ref{eq:Ap}.  The E866 group is 
thus able to measure the ratio $\overline{d}^p/\overline{u}^p$ over a wide
range of $x$ \cite{e866}.

\subsection{Drell-Yan Asymmetry for Kaon$^\pm$ Beams}

For $K^+$ projectile we have
\begin{equation}
H_{K^+p}^{DY} = {1 \over 9}\left[4\left( 
  u^{K^+}_v\,\bar{u}^p_s + \bar{u}^{K^+}_s\,u^p_v \right) + 
  \bar{d}^{K^+}_s\,d^p_v + \bar{s}^{K^+}_v\,s^p_s \right],
\label{eq:Kpp}
\end{equation}
and a similar expression for $H^{DY}_{K^+n}$. Here we have suppressed
for convenience the dependence of the parton distributions on $x_1$
or $x_2$, since there is no ambiguity in this case. The asymmetry is
\begin{eqnarray}
A^{DY}_{K^+} & = & {H_{K^+p}^{DY}- H_{K^+n}^{DY}\over 
  H_{K^+p}^{DY}+ H_{K^+n}^{DY}} \nonumber \\ 
  &=& {-4u^{K^+}_v\left( \bar{d}^p_s - 
  \bar{u}^p_s \right) + \left( 4\bar{u}^{K^+}_s - \bar{d}^{K^+}_s 
  \right)\left(u^p_v - d^p_v\right)\over 
  4u^{K^+}_v\left( \bar{d}^p_s + 
  \bar{u}^p_s \right) + \left( 4\bar{u}^{K^+}_s + \bar{d}^{K^+}_s 
  \right)\left(u^p_v + d^p_v\right) + 2\bar{s}^{K^+}_v\,s^p_s}. 
\label{eq:AKpp}
\end{eqnarray}

In Fig.\ \ref{fig3} we show the asymmetry $A^{DY}_{K^+}(x_1=x_2)$ vs.\
$x_1$. The curves use the GRV parton distributions for the nucleons
\cite{grv}. Since little experimental information is available on kaon
structure functions, we use in their place the GRV distributions for
the pion \cite{grv}. The lower curve uses the full parton
distributions, while the upper curve is obtained if we fix
$\bar{u}^p(x) = \bar{d}^p(x)$.  Note that for $x$ between 0.1 and 0.4,
the difference between the two asymmetries is quite large. If one had
an intense well separated $K^+$ beam, it would, in principle, be an
excellent probe of flavor asymmetry in the proton sea.

As is clear from Eq.\ \ref{eq:AKpp}, in order to calculate 
$A^{DY}_{K^+}$ we need {\it four} different parton distribution
functions for $K^+$, while the only experimental information we
have is on the ratio $\bar{u}^{K^-}/\bar{u}^{\pi^-}$ \cite{Bad80}.  
This ratio was found to be consistent with unity for $x_1 \leq 0.7$, 
and was less than unity only for larger values of $x_1$.  It
was determined by using an unseparated negative beam containing
pions, kaons and antiprotons, and by making the following
assumptions: the meson sea was neglected, among the valence 
quarks in the mesons only the {\it up} flavor was retained, and 
finally the Drell-Yan scale factors for $K^-$ and $\pi^-$ were
assumed to be identical and constant over the kinematic range
explored.  This points out the urgent need for more experimental
data on kaon structure functions, and justifies our use of
pion structure functions in place of kaon structure functions 
in the calculation of the Drell-Yan asymmetry.  

We again examine whether the region of equal-$x$ kinematics is the
most sensitive region to probe the flavor asymmetry in the proton sea.
Figures 4 a-c correspond to Figs. 2 a-c, respectively, with the projectile
proton replaced by a $K^+$ projectile. In the case of the proton
projectile, we saw that the equal-$x$ kinematics was adequate for the
purpose of maximizing the difference $\Delta^{DY}_p$ (Fig. 2c), but
was not optimum for the purpose of maximizing the asymmetry $A^{DY}_p$
(Fig. 2a). {\it In the case of the $K^+$ projectile, we find that the
equal-$x$ kinematics is not good for either purpose} (Figs. 4a and 4c).
It is clear from Fig. 4a that if $x_1 = x_2 < 0.3$ the asymmetry is
nearly zero, but can be significantly enhanced by deviating from the
$x_1 = x_2$ line.

The asymmetry for $K^-$ mesons incident on protons and neutrons (in
the deuterium) can be calculated similarly. Using the same
approximations as before, we obtain
\begin{equation}
H_{K^-p}^{DY} = {1 \over 9}\left[4\left( 
  \bar{u}^{K^-}_v\,u^p_v + \bar{u}^{K^-}_s\,u^p_v + 
  \bar{u}^{K^-}_v\,u^p_s \right) + 
  \bar{d}^{K^-}_s\,d^p_v + s^{K^-}_v\,\bar{s}^p_s \right],
\label{eq:Kmp}
\end{equation}
and a similar expression for $H^{DY}_{K^- n}$. The asymmetry is
\begin{eqnarray}
A^{DY}_{K^-} & = & {H_{K^-p}^{DY}- H_{K^-n}^{DY}\over 
  H_{K^-p}^{DY}+ H_{K^-n}^{DY}} \nonumber \\ 
  &=& { -4\bar{u}^{K^-}_v \left(\bar{d}^p_s - \bar{u}^p_s \right) 
  + \left(4\bar{u}^{K^-}_v + 4\bar{u}^{K^-}_s - 
  \bar{d}^{K^-}_s \right)\left( u^p_v - 
  d^p_v \right)\over
  4\bar{u}^{K^-}_v \left(\bar{d}^p_s + \bar{u}^p_s\right) +
  \left( 4\bar{u}^{K^-}_v + 4\bar{u}^{K^-}_s + \bar{d}^{K^-}_s
  \right)\left( u^p_v + d^p_v \right) + 2s^{K^-}_v \bar{s}^p_s}.
\label{eq:AKmp}
\end{eqnarray}
Numerical calculations show that this asymmetry is insensitive
to the up/down difference in the proton sea-quark distributions
(see section IV).

\subsection{Drell-Yan Asymmetry for Pion$^\pm$ Beams}

For $\pi^+$ projectile we have
\begin{equation}
H_{\pi^+p}^{DY} ={1 \over 9}\left[4\left( 
  u^{\pi^+}_v\,\bar{u}^p_s + \bar{u}^{\pi^+}_s\,u^p_v \right) + 
  \bar{d}^{\pi^+}_v\,d^p_v + \bar{d}^{\pi^+}_v\,d^p_s 
  + \bar{d}^{\pi^+}_s\,d^p_v \right],
\label{eq:Pipp}
\end{equation}
and a similar expression for $H_{\pi^+n}^{DY}$. The asymmetry is
\begin{eqnarray}
A^{DY}_{\pi^+} & = & {H_{\pi^+ p}^{DY}- H_{\pi^+n}^{DY}\over 
  H_{\pi^+p}^{DY}+ H_{\pi^+n}^{DY}} \nonumber \\ 
  &=& {- \left( 4u^{\pi^+}_v - \bar{d}^{\pi^+}_v \right)\left( 
  \bar{d}^p_s - \bar{u}^p_s \right) + \left( 4\bar{u}^{\pi^+}_s 
  - \bar{d}^{\pi^+}_v - \bar{d}^{\pi^+}_s \right) \left( 
  u^p_v - d^p_v\right)\over 
  \left( 4u^{\pi^+}_v + \bar{d}^{\pi^+}_v \right)\left( 
  \bar{d}^p_s + \bar{u}^p_s \right) + \left( 4\bar{u}^{\pi^+}_s 
  + \bar{d}^{\pi^+}_v + \bar{d}^{\pi^+}_s \right) \left( 
  u^p_v + d^p_v\right) }.
\label{eq:APipp}
\end{eqnarray}
For $\pi^-$ beams, one obtains
\begin{equation}
H_{\pi^-p}^{DY} ={1 \over 9}\left[4\left( 
  \bar{u}^{\pi^-}_v\,u^p_v + \bar{u}^{\pi^-}_s\,u^p_v + 
  \bar{u}^{\pi^-}_v\,u^p_s \right) + 
  d^{\pi^-}_v\,\bar{d}^p_s + \bar{d}^{\pi^-}_s\,d^p_v \right],
\label{eq:Pimp}
\end{equation}
and a similar expression for $H_{\pi^-n}^{DY}$. The asymmetry is
\begin{eqnarray}
A^{DY}_{\pi^-} & = & {H_{\pi^- p}^{DY}- H_{\pi^-n}^{DY}\over 
  H_{\pi^-p}^{DY}+ H_{\pi^-n}^{DY}} \nonumber \\ 
  &=& {-\left(4\bar{u}^{\pi^-}_v - d^{\pi^-}_v \right) \left( 
  \bar{d}^p_s - \bar{u}^p_s \right) +
  \left( 4\bar{u}^{\pi^-}_v + 4\bar{u}^{\pi^-}_s - 
  \bar{d}^{\pi^-}_s \right)\left( u^p_v - d^p_v \right) \over 
 \left( 4\bar{u}^{\pi^-}_v + d^{\pi^-}_v \right)\left( 
  \bar{d}^p_s + \bar{u}^p_s \right) + \left( 4\bar{u}^{\pi^-}_v 
  + 4\bar{u}^{\pi^-}_s + \bar{d}^{\pi^-}_s \right) \left( 
  u^p_v + d^p_v \right) }.
\label{eq:APimp}
\end{eqnarray}
Numerical calculations show that both $A^{DY}_{\pi^+}$ and
$A^{DY}_{\pi^-}$ are insensitive to the up/down difference in the 
proton sea-quark distributions (see section IV).

\section{Drell-Yan Sum Rules}

If one assumes (a) charge conjugation symmetry so that e.g.,
$q^{\pi^+} = \bar{q}^{\pi^-}$, (b) isospin symmetry so that e.g.,
$u^{\pi^+} = d^{\pi^-}$, (c) $q_{sea}(x) = \bar{q}_{sea}(x)$, and (d)
that charm and heavier flavors are negligible, then it is
straightforward to show that $\pi^+$ has only three charged-parton
distribution functions (PDF), namely
\begin{eqnarray}
 u_{val}(x) &=& \bar{d}_{val}(x) \equiv V^{\pi}(x), \nonumber \\
 u_{sea}(x) &=& \bar{u}_{sea}(x) = d_{sea}(x) = \bar{d}_{sea}(x) 
\equiv S^{\pi}_{light}(x), \nonumber \\
 s_{sea}(x) &=& \bar{s}_{sea}(x) \equiv S^{\pi}_{heavy}(x),
\end{eqnarray}
and those for $\pi^-$ can be identified with one of these three. Thus
$\pi^+$ and $\pi^-$ together have only three charged-parton
distribution functions. With the same set of assumptions (a)-(d), it
is easy to show that $K^+,\ K^-,\ K^0$ and $\bar{K}^0$ together have
only five charged-parton distribution functions which we denote by
\begin{equation}
V^K_{light}(x),~~~ V^K_{heavy}(x),~~~ S^K_{light,1}(x),~~~ 
S^K_{light,2}(x),~~~ S^K_{heavy}(x),
\end{equation}
where the notation is as in Eq. (3.1).

To experimentally determine the three PDF's for pions, Eq. (3.1), one
needs data on any three of the following four quantities:
\begin{equation}
H^{DY}_{\pi^- D},~~~ H^{DY}_{\pi^- p},~~~ H^{DY}_{\pi^+ D},~~~
 H^{DY}_{\pi^+ p}.
\end{equation}
Out of these, the first three may be preferable due to higher event
rates. In fact, if the sea-sea contribution to $H^{DY}_{\pi}$ is
ignored then $H^{DY}_{\pi}$ would contain only $V^{\pi}(x)$ and
$S^{\pi}_{light}(x)$, and data on only two of the four quantities in
(3.3) would be sufficient.

Similarly, if the sea-sea contribution to $H^{DY}_K$ is ignored, then
$S^K_{heavy}(x)$ does not occur in $H^{DY}_K$. To determine the
remaining four PDF's for $K$, one needs data on
\begin{equation}
H^{DY}_{K^- D},~~~ H^{DY}_{K^- p},~~~ H^{DY}_{K^+ D},~~~
H^{DY}_{K^+ p}.
\end{equation}
If further the difference between $S^K_{light,1}$ and $S^K_{light,2}$
is ignored, then the data on any three quantities in (3.4)
would be sufficient.

As an alternative, some of these PDF's can be determined with the help
of DY sum rules which we now present. For a beam particle $B$ incident
on the deuterium ($D$) target, we assume $H^{DY}_{BD} = H^{DY}_{Bp} +
H^{DY}_{Bn}$. Recall also that $x_1 \equiv x_{beam}$ and $x_2 \equiv
x_{target}$. It is straightforward to show that
\begin{eqnarray}
\int ~~(H^{DY}_{\pi^-p} - H^{DY}_{\pi^+p} ) ~~dx_2 &=& 7/9 
~.~ V^{\pi}(x_1),
\nonumber \\
\int ~~(H^{DY}_{\pi^-D} - H^{DY}_{\pi^+D} ) ~~dx_2 &=& V^{\pi}(x_1),
\nonumber \\
\int ~~(H^{DY}_{K^-p} - H^{DY}_{K^+p} ) ~~dx_2 &=& 8/9 
~.~ V^{K}_{light}(x_1),
\nonumber \\
\int ~~(H^{DY}_{K^-D} - H^{DY}_{K^+D} ) ~~dx_2 &=& 4/3 
~.~ V^{K}_{light}(x_1),
\label{sr1}
\end{eqnarray}
where the limits of integration are 0 and 1. Note the interesting fact
that these rules do not require any knowledge of the target
particle structure functions. If the integration was done over $x_1$
instead of $x_2$, we would have got a different set of sum rules with
the well known nucleon structure functions, instead of the poorly
known meson structure functions, occurring on the right-hand sides of
Eqs. (\ref{sr1}). 

There are additional sum rules where the integration is done over both
$x_1$ and $x_2$:
\begin{eqnarray}
\int \int ~~(H^{DY}_{\bar p p} - H^{DY}_{p p} ) ~~dx_1 ~~dx_2 &=& 17/9,
\nonumber \\
\int \int ~~(H^{DY}_{\bar p D} - H^{DY}_{p D} ) ~~dx_1 ~~dx_2 &=& 3,
\nonumber \\
\int \int ~~(H^{DY}_{K^- p} - H^{DY}_{K^+ p} ) ~~dx_1 ~~dx_2 &=& 8/9,
\nonumber \\
\int \int ~~(H^{DY}_{K^- D} - H^{DY}_{K^+ D} ) ~~dx_1 ~~dx_2 &=& 4/3,
\nonumber \\
\int \int ~~(H^{DY}_{\pi^- p} - H^{DY}_{\pi^+ p} ) ~~dx_1 ~~dx_2 &=& 7/9,
\nonumber \\
\int \int ~~(H^{DY}_{\pi^- D} - H^{DY}_{\pi^+ D} ) ~~dx_1 ~~dx_2 &=& 1.
\label{sr2}
\end{eqnarray}
If QCD corrections are included, all the right-hand sides in the above
sum rules will get multiplied by the well known DY $K$ factor whose
value is about 2. Since the nucleon PDF's are rather well known, one
may also look at the first of Eqs. (\ref{sr2}) as a way to determine
the average DY $K$ factor. It may not be easy to verify these sum
rules experimentally, because each rule involves two experiments
(e.g., with $\pi^-$ and $\pi^+$ beams). Moreover, the integration is
over the entire kinematic range which would entail some extrapolation
of the data. However, if the beams do not have great intensity, the
integration may help in getting enough counts to make a measurement.
Thus, if suitable beams are available, experimental verification may
not be impossible. Sum rules similar to those in Eq. (\ref{sr2}) were
given by Hwa \cite{hwa}. The fact that the valence contributions to
the DY process can be isolated by taking the difference between
particle beam and antiparticle beam on a given target was mentioned by
Sarma \cite{sarma}, but no sum rule was given.

\section{Conclusions}

{\it Why are the proton and $K^+$ projectiles more sensitive than the
$K^-,\ \pi^+$ and $\pi^-$ projectiles, to the flavor asymmetry in the
proton sea?} Inspection of the expressions for $A^{DY}_B$ for various
projectiles shows that they all contain a contribution proportional to
$( \bar{d}^p_s(x_2) - \bar{u}^p_s(x_2) )$ and a
contribution proportional to $\left( u^p_v(x_2) - d^p_v(x_2)
\right)$. The former constitutes the ``signal'' while the latter is 
``noise''. For all five projectiles, the signal is getting
multiplied by valence parton distribution functions in the projectile.
But these projectiles differ from each other in the following respect.
In the case of $p$ and $K^+$ projectiles, the noise is getting
multiplied by the {\it sea} parton distribution functions from the
projectile, while for the $K^-,\ \pi^+$ and $\pi^-$ projectiles, the
noise is getting multiplied by {\it valence} parton distribution
functions from the projectile. Naturally in the latter case, the
signal is expected to be relatively weak. This is borne out by the
numerical calculations presented here. The difference between the
numerical results for $p$ and $K^+$ projectiles on one hand and the
$K^-,\ \pi^+$ and $\pi^-$ projectiles on the other, can be traced to
the fact that $p$ and $K^+$ contain neither $\bar{u}$ nor $\bar{d}$ as
valence partons, while $K^-,\ \pi^+$ and $\pi^-$ do.

Results presented in Figs. 2 and 4 bring into focus the kinematic
regions which can be optimally explored in future experiments designed
to probe the flavor asymmetry in the proton sea. Finally we have
presented sets of Drell-Yan sum rules; those in Eq. (3.5), for
example, could throw some light on the poorly known meson structure
functions. The first of Eqs. (3.6) could be used to deduce the DY $K$
factor averaged over the full kinematic region.


\acknowledgments

One of the authors [RSB] thanks the Indiana University Nuclear Theory
Center for its hospitality during the period this work was undertaken.
One of the authors [JTL] was supported in part by the US NSF under
research contracts NSF-PHY-9408843 and NSF-PHY-9722076. RSB thanks the
late Prof. K.V.L. Sarma for many useful discussions and comments on
the manuscript.



\begin{figure}
\caption{
Theoretical Drell-Yan asymmetry $A^{DY}_p(x_1=x_2)$ vs.\ $x_1$ for
proton beam; see Eq. (2.5). Here $x_1 \equiv x_{beam}$ and $x_2 \equiv 
x_{target}$. The lower curve is calculated with full parton
distributions, the upper one fixing $\bar{u}^p(x) = \bar{d}^p(x)$.} 
\label{fig1}
\end{figure}    

\begin{figure}
\caption{
Contour plots of Drell-Yan asymmetries for the proton beam. (a)
Contours of equal asymmetry $A^{DY}_p$, using full GRV parton
distributions of Ref. 
[9]. 
(b) Contours of equal asymmetry
using GRV parton distributions but fixing $\bar{u}^p(x) =
\bar{d}^p(x)$. (c) Contours of equal difference $\Delta^{DY}_p$,
defined in Eq. 
(2.6). 
The two extreme contours are labelled;
the intermediate contours are understood to have labels differing in
steps of 0.1. Dashed curve: $x_{beam} = x_{target}$.}  
\label{fig2}
\end{figure}

\begin{figure}
\caption{
Same as Fig. 1 except that the proton beam is replaced by the $K^+$
beam.}
\label{fig3}
\end{figure}    

\begin{figure}
\caption{
Same as Fig. 2 except that the proton beam is replaced by the $K^+$
beam.}
\label{fig4}
\end{figure}    

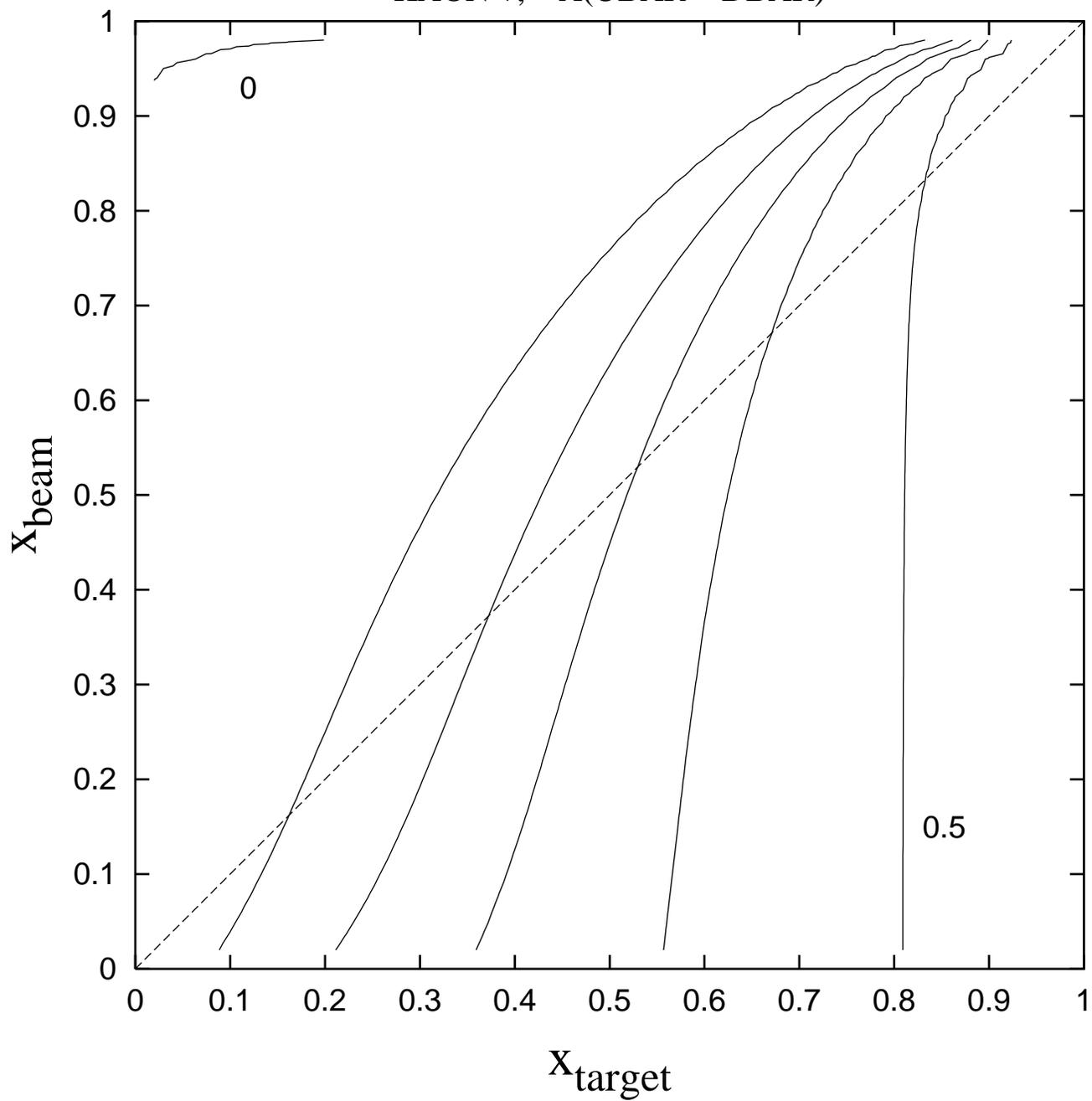
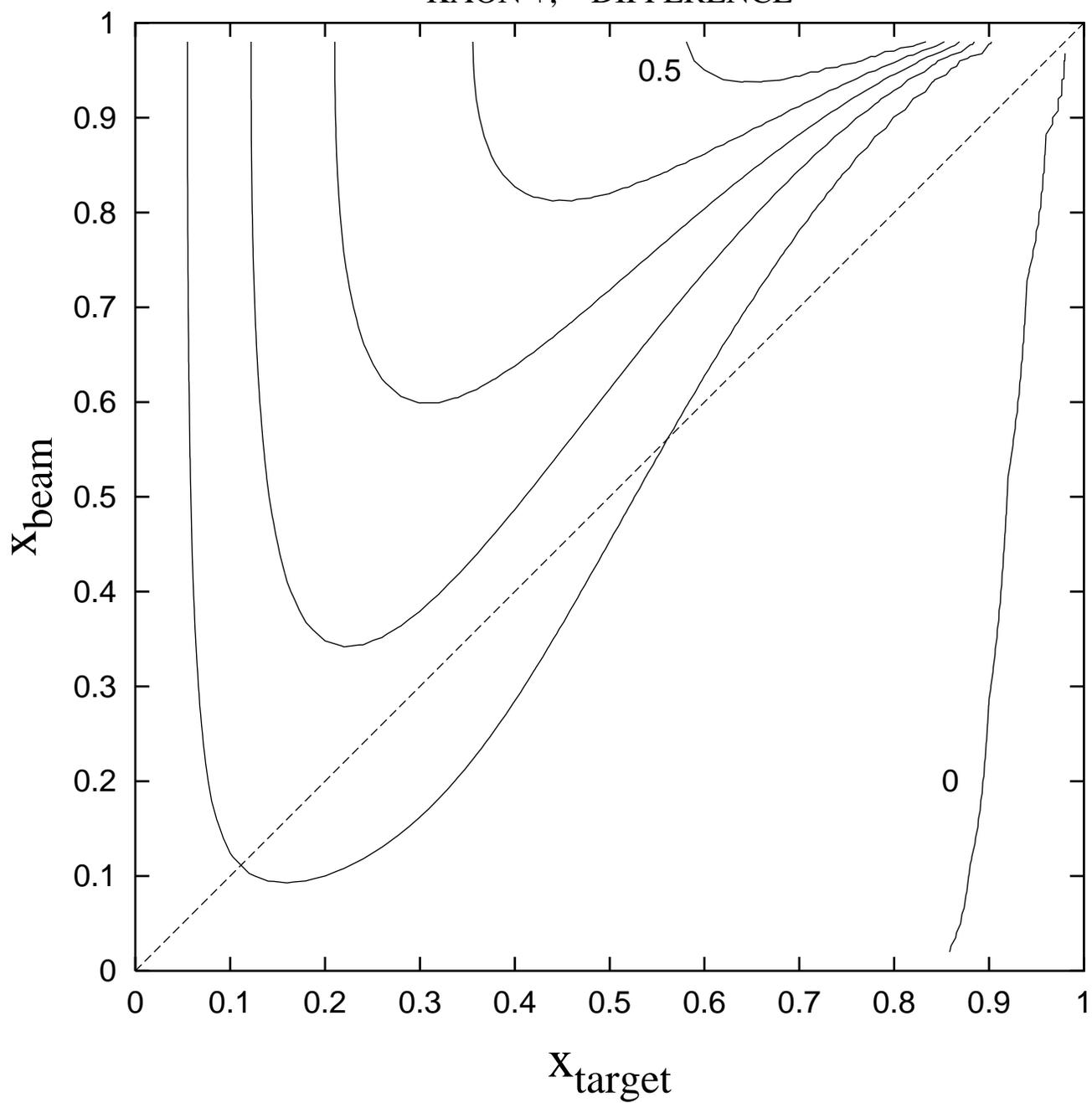
\end{document}